# Effect of Weighting Scheme to QoS Properties in Web Service Discovery

[1]Agushaka J. O., Lawal M. M., Bagiwa, A. M. and Abdullahi B. F.

Mathematics Department, Ahmadu Bello University Zaria-Nigeria

[1]jagushaka@yahoo.com

## Abstract

Specifying QoS properties can limit the selection of some good web services that the user will have considered; this is because the algorithm used strictly ensures that there is a match between QoS properties of the consumer with that of the available services. This is to say that, a situation may arise that some services might not have all that the user specifies but are rated high in those they have. With some tradeoffs specified in form of weight, these services will be made available to the user for consideration. This assertion is from the fact that, the user's requirements for the specified QoS properties are of varying degree i.e. he will always prefer one ahead of the other. This can be captured in form of weight i.e. the one preferred most will have the highest weight. If a consumer specifies light weight for those QoS properties that a web service is deficient in and high weight for those it has, this will minimize the difference between them. Hence the service can be returned.

## 1. Introduction

Web Services are the third generation web applications; they are modular, self-describing, self-contained applications that are accessible over the Internet Cubera et al (2001). A Web Services (sometimes called an XML Web Services) is an application that enables distributed computing by allowing one machine to call methods on other machines via common data formats and protocols such as XML and HTTP. Web Services are accessed, typically, without human intervention. Web service technology address the problem of platform interoperability however, in the work of Plammer and Andrews (2001), they showed that there is actually a slow take off of web services technology and DuWaldt and Trees

(2002) attributed this slow take off to factors such as perceived lack of security and transaction support and also quality of the web service. Web Services standards like WSDL (www.w3.org/TR/wsdl), SOAP (www.w3.org/TR/soap2-part1), UDDI (www.uddi.org/pubs/uddi-v3.00) and BPEL (ftp://www6.software.ibm.com/software/developer/library/ws-bpel.pdf) provide syntax based interaction and composition of Web Services in a loosely coupled way that does not take into account the non-functional specification like quality of service (QoS) properties such as scalability, performance, accessibility etc. QoS for Web services gives consumers assurance and confidence to use the services, consumers aim to experience a good service performance, e.g. low waiting time, high reliability, and availability to successfully use services. Service registries host hundreds of similar Web services, which make it difficult for the service consumers to choose from, as the selection is only based on the functional properties albeit they differ in QoS that they deliver. Such variety in QoS is considered as an important criterion for Web service selection. Taher, L. et al (2005a) proposed a generic QoS Information and Computation (QoS_IC) framework for QoS-based service selection in which the QoS selection mechanism utilizes an established *Registry Ontology:* which is used to present the semantics of the proposed framework and its QoS structure. The QoS selection mechanism also uses the Euclidian distance measure to evaluate the similarity between the consumer/provider QoS specification in the matchmaking process. We try to extend the work of Taher et al (2005a) to accommodate a user defined weighting scheme. This weighting scheme is defined in such a way that the highest weight signifies the most desired QoS property. It decreases base on order of priority. Also, the weighting scheme normally between [0,1]. The algorithm presented here is a slight modification of Taher's as it take into consideration the weighting scheme. As part of the aim of this paper, we show that the introduction of a weighting scheme into the discovery algorithm can greatly address the issues of "trade off" that can arise in service selection. That is, depending on the weight specification, certain web services can perform better and hence be returned. The examples in this paper helped us in making these assertions. The sections in this paper are organized as follows: Related work is given next, it is closely followed by QoS matching in Tahers work, then our propose extension. Detailed examples are given next to proof our assertions. Finally, conclusion and future work is given.

## 2. Related work

At the present time, Universal Description, Discovery and Integration of Web services (UDDI) based look ups for Web services are based on the functional aspects of the desired Web services. In his work, Ran (2003) extended UDDI model by adding a new role called QoS certifier which verifies the service providers QoS claims. Figure 1 is an adaptation from the work of Ran (2003). In his proposed model, Ran assumes that a Web service provider needs to supply information about the company, the functional aspects of the provided service as requested by the current UDDI registry, as well as to supply quality of service information related to the proposed Web service. The claimed quality of service needs to be certified and registered in the repository.

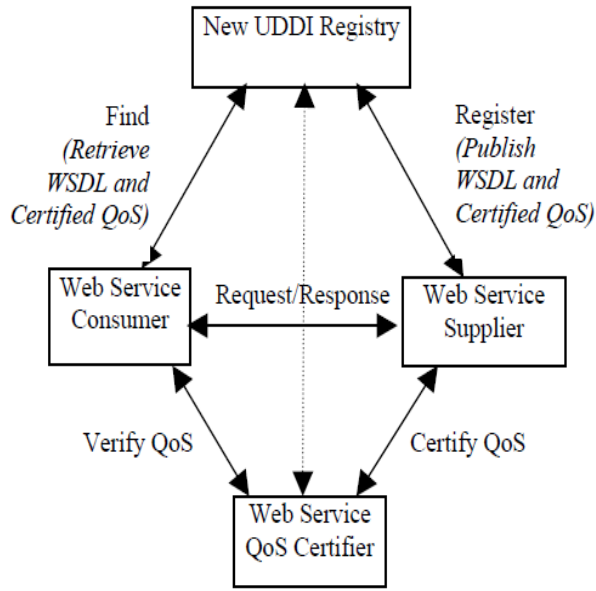

Figure 1: Rans UDDI model

The consumer searches the UDDI registry for a Web service with the required functionality as usual; they can also add constraints to the search operation. One type of constraint is the required quality of service. If there were multiple Web services in the UDDI registry with similar functionalities, then the quality of service requirement would enforce a finer search. The search would return a Web service that offers the required functionality with the desired set of quality of service. If there is no Web service with these qualities, feedback is given to the consumer. This approach lacks support for the dynamic nature of these QoS properties. The approach of Taher et al (2005a) takes into account this issue of dynamic nature of QoS properties. We implemented his work using a different similarity metric and also improve his matching algorithm to accommodated user defined weighting scheme for the QoS properties. Depending on which property the user desires best he gives it a higher weight. Apart from this two works, several attempt have been made to add QoS specification to the discovery process of web services. Examples of such approaches are: Web Service Level Agreement (WSLA) (Keller, A. et al, 2002), Web service-QoS (Ws-QoS) (Tian, M., 2004), Web Service Offering Language (WSOL) (Tosic, V., 2003), SLAng (Lamanna, D., 2003), UDDI eXtension (UX) (Chen, Z., 2003) and UDDIe (Ali, S., 2003). A comparison of the work of Taher et al (2005a) and other approaches is given in (Taher et al, 2005b). This serves as basis for our selection of Taher's work for improvement. All such approaches do not address issues related to adapting the consumers to the changing conditions of providers systems (Taher et al, 2005a).

## 3. QoS Matching in Taher's Work

Matchmaking problem meets the question of distance measure between objects, there are many approaches to measure distance between any two objects based on their numerical or semantic closeness, the Euclidean distance measure was chosen for the algorithm. In other words, Euclidean distance is used to evaluate the square root of the sum of squared differences between corresponding elements of the two vectors

### 3.1. The QoS matchmaking algorithm

The QoS matchmaking algorithm determines which Web service $ws_i$, from WS, $WS = \{ws_1, ws_2, ws_3, \ldots ws_n\}$, is selected based on consumer's QoS specifications (QPc). For that purpose, QoS Manager constructs $QoS_{nk}$ matrix, where n represents the total number of Web services (WS) that have the same functional properties, and k represent the total number of QoS properties. To compensate between different measurement units between different QoS properties values ($qp_{i,v}$), the values need to be normalized to be in the range [0, 1]. We will use the following equations to normalize them.

$$Norm(qp_{i,v}) = \frac{qp_{i,v_{max}} - qp_{i,v}}{qp_{i,v_{max}} - qp_{i,v_{min}}} \qquad 1$$

$$Norm(qp_{i,v}) = \frac{qp_{i,v} - qp_{i,v_{min}}}{qp_{i,v_{max}} - qp_{i,v_{min}}} \qquad 2$$

Where $qp_{i,v}$, is the QoS property that one wishes to normalize by minimization using equation-1 or maximization using equation-2, for example, response time need to be normalized by minimization using equation-1 while availability needs to be normalized by maximization using equation-2. $qp_{i,v_{max}}$ is the $qp_{i,v}$ that has the maximum value among all values on column v and $qp_{i,v_{min}}$ is the $qp_{i,v}$ that has the minimum value among all values on column v. To normalize matrix QoS, we need to define an array $NR$, $NR = \{nr_1, nr_2, \ldots \ldots nr_k\}$ with $1 \leq v \leq k$. The value of $nr_v$ can be either 0 or 1, 0 indicates that $qp_v$ should be normalized using equation-1, whereas 1 indicates that $qp_v$ should normalized using equation-2. The key idea of the QoS matchmaking algorithm is to find the nearest $ws_i$ to the QoS specifications of the consumer (QPc), i.e. to find a Web service with a minimum Euclidian distance. Given the QoS mode (qm) and the submitted QoS properties (QPc) submitted by the consumer, QoS matchmaking algorithm works as follows:

**Step-1**: Check qm, based on that

**Step-2**: Construct QoS matrix.

**Step-3** Normalizes (QPc) using equation-1 and equation-2.

**Step-4-5** Normalize QoS matrix using equation-1 and equation-2.

**Step-6-7** Compute the Euclidian distance between each QP(wsi) and (QPc).

**Step-7** Find $ws_i$ with the minimum distance

## 4. Proposed Extensions

In this section, we give the extensions proposed for the model given in Taher et al (2005a). The assumption here is that all other components given in Taher's model remain except the similarity metric and the QoS matching algorithm. Detail is given in the following sections:

### 4.1. Similarity Metric

In their work, Taher et al (2005a) used the Euclidean distance to measure the similarity between two vectors. This does not capture any form of weighting for the QoS properties. As an extension to their work, we introduced a user specified weighting vector $U_{weight_i}$. As we have said earlier is normally between [0,1]. The modified formula is given below:

- Euclidean distance measure is used to evaluate the similarity between two vectors $t_i = t_{i1} \ldots \ldots t_{ik}$ and $t_j = t_{j1} \ldots \ldots t_{jk}$. Here we introduce $U_{weight_i}$ which we use to multiply the Euclidean distance.

$$dis(t_i, t_j) = \sqrt{\sum_{h=1}^{k}(t_{ih} - t_{jh})^2 \times U_{weight_i}}$$

## 4.2. Assertion

We are saying that the introduction of a weighting scheme will help our algorithm to accommodate the tradeoffs that exists in nature. Specifying QoS properties can limit the selection of some good web services that the user will have considered, as the algorithm strictly ensures that there is a match between QoS properties of the consumer with that of the available services. This is to say that, some services might not have all that the user specifies but are rated high in those they have. With some tradeoffs specified in form of weight, these services will be made available to the user for consideration. This assertion is from the fact that, the user's requirements for the specified QoS properties are of varying degree i.e. he will always prefer one ahead of the other. This can be captured in form of weight i.e. the one preferred most will have the highest weight. If a consumer specifies light weight for those QoS properties that a web service is deficient in and high weight for those it has, this will minimize the difference between them. Hence the service can be returned as it shows from case 2.

We will use the same example given in Taher et al (2005a) to proof our assertion

## 5. The new algorithm

This new algorithm is an improved version of that given in Taher et al (2005a). This is because it incorporates a user defined weighting scheme for the desired QoS properties.

Given web services $WS = \{ws_1, ws_2, \ldots \ldots ws_n\}$ that satisfy the user's functional requirements, QoS properties $QP_c$ and QoS properties weight base on the user's specified priority. Just as in the work of Taher et al (2005a), this algorithm tries to find which of the web services $ws_i$ that best satisfy the consumer's request based on the non-functional specification (QoS). For this purpose, a quality matrix, $\mathbb{Q} = \{V(Q_{ij}); 1 \leq i \leq m; 1 \leq j \leq n\}$ is created, this refers to a collection of quality attribute-values for a set of candidate services, such that, each row of the matrix corresponds to the value of a particular QoS attribute (in which the user is interested) and each column refers to a particular candidate service. In other words, $V(Q_{ij})$, represents the value of the $i^{th}$ QoS attribute for the $j^{th}$ candidate service. The normalization equations given in Taher's work is used to normalize QoS properties obtained from profile of web services and $QP_c$ to be in the range [0,1]. Given the QoS mode (qm) as in Taher et al (2005a), the submitted QoS properties $(QP_c)$ submitted by the consumer and the QoS weight, QoS matchmaking algorithm works as follows:

**Step-1:** Check qm, based on that

**Step-2:** Construct Quality matrix.

**Step-3:** Normalize ($QP_c$) and quality matrix.

**Step-4:** Compute the similarity using the metric given in previous section, between each $QP(WS_i)$ and ($QP_c$).

**Step-5** Find $ws_i$ with the minimum distance

Just as in the case of Taher's alogorithm, The QoS matchmaking algorithm works fine even if ($QP_c$) does not include the whole set of QoS properties, as it is anticipated that consumers need not to specify all QoS parameters defined previously. The Complexity is **O(n),** since the number of QoS parameters is constant and n represents the total number of Web services that have the same functionality based on the consumer's functional requirements. However the complexity could change, in case the number of QoS properties change to a large value.

## 6. Example

This example is adopted from Taher et al (2005a). It considers a scenario of how the QoS matchmaking algorithm works.

Assume that $QP_c = \{0.9, 20, 50, 0.9, 1, 200\}$, $qm = (WHM/NTM)$ and $NR = \{1, 0, 1, 1, 1, 0\}$, the QoS properties are in order of scalability, response time, throughput, availability, accessibility and cost. Also, assuming the user specify weighting schemes for the QoS properties as follows

1. $U_{weight_i} = \{0.9, 1, 0.6, 0.4, 0.6, 0.1\}$
2. $U_{weight_i} = \{0.9, 0.1, 1, 0.1, 0.2, 0.9\}$

In these schemes, two different tradeoffs or variation in user's wants are shown. Based on the functional specifications assume that four Web services $\{ws_1, ws_2, ws_3, ws_4\}$ have been returned by UDDI. QoS matchmaking algorithm retrieves the relevant QoS properties associated with $qm$ mode of the four Web services and use it to construct Quality matrix, as shown in table-1.

Table-1: quality matrix

|       | $qp_{sc}$ | $qp_{res}$ | $qp_{Thr}$ | $qp_{avl}$ | $qp_{acs}$ | $qp_{cost}$ |
|-------|------|------|------|------|------|------|
| $ws_1$ | 0.9  | 10   | 100  | 1    | 0.9  | 500  |
| $ws_2$ | 0    | 15   | 30   | 0.8  | 0.6  | 100  |
| $ws_3$ | 0.3  | 5    | 20   | 0.6  | 0.9  | 200  |
| $ws_4$ | 1    | 20   | 200  | 0.9  | 1    | 300  |

QoS matchmaking algorithm continues by normalizing $QP_c$ and Quality matrix. The QoS values of $QP_c$ after normalization are $\{0.90, 0.00, 0.17, 0.75, 1.00, 0.75\}$. The QoS values of Quality matrix after normalization are shown in table-2

Table-2: normalized QoS matrix

|       | $qp_x$ | $qp_{res}$ | $qp_{Thr}$ | $qp_{avl}$ | $qp_{acs}$ | $qp_{cost}$ |
|-------|------|------|------|------|------|------|
| $ws_1$ | 0.90 | 0.67 | 0.44 | 1.00 | 0.75 | 0.00 |
| $ws_2$ | 0.00 | 0.33 | 0.06 | 0.50 | 0.00 | 1.00 |
| $ws_3$ | 0.30 | 1.00 | 0.00 | 0.00 | 0.75 | 0.75 |
| $ws_4$ | 1.00 | 0.00 | 1.00 | 0.75 | 1.00 | 0.50 |

The algorithm then calculates the similarities for the web services by using the our extended formula given earlier

1. Case 1: using $U_{weight_i} = \{0.9, 1, 0.6, 0.4, 0.6, 0.1\}$

The algorithm calculates the distance for $ws_1, ws_2, ws_3$ and $ws_4$ to be $0.782, 1.215, 1.266$ and $0.655$ respectively. Since $ws_4$ has the minimum, it is returned.

Note: comparing our result with that in Taher et al (2005a), we see that the distance in all the web services is greatly reduced. Remarkably, if look at the distance for $ws_1$, as compared to that in Taher's work we see that the distance is significantly reduced. This is because the user is less interested in QoS property cost as shown by its weight, in which the web service differ from the request ( as seen from both the QoS profile of the service and request)

2. Case 2: using $U_{weight_i} = \{0.9, 0.1, 1, 0.1, 0.2, 0.9\}$

The algorithm returns distances for $ws_1, ws_2, ws_3, ws_4$ to be $0.8017, 1.00, 0.7222$ and $0.8684$.

Since $ws_3$ has the minimum, it is returned.

Note: there is a significant decrease in distance for $ws_3$ this is because the user is less interested in QoS properties, as indicated by their weights, which $ws_3$ is not rated high for. As compared to case 1 where $ws_4$ was returned with the least difference, we see that the weighting scheme for case 1 has more weight in areas where the similarities are high between $ws_4$ and user required QoS properties. This goes to show the effect of weighting scheme in determining which service is returned. Though the consumer wants all the QoS properties he

specified, tradeoffs specified in form of weights helps return services which hitherto will not be considered because of their low weight in some QoS properties which will have caused their difference to increase. This case confirms our assertion that weight plays a great deal in returning services that actually meets the consumer's needs. The fact is that there is always trade off that exists in these needs and these tradeoffs are best captured using a weighting scheme.

As shown in the above example the results are promising and proofing our concept. In our ongoing work, we are attempting to provide an implementation for this work. This we hope will be based on the implementation of Taher's work.

## 7. Conclusion

Quality of Service selection for Web services is becoming a significant challenge. We proposed an advanced QoS based selection framework that is an improvement on the work Taher et al (2005a) that manages Web service quality and provides mechanisms for QoS updates. The proposed improvement preserves the architecture presented in Taher's work and proposes an extension that includes a user defined weighting scheme in both metric used to calculate similarities and the matching algorithm. This extension can be added seamlessly without any change in the architecture in Taher's work, it also can be customized for specific domain. From the example given herein especially in case 2, we see that adding a weighting scheme to the QoS matching process greatly affects the service to be return. Low weight to QoS property a web service has low rating in will make it insignificant and greatly affects its (service) selection as can be seen in case 2 of our example. Our work also works base on the assumption in Taher's work that the QoS Modes have been derived with the assumption that network conditions are static; in practice network factors would have a direct affect on many QoS properties such as response time, we are working on addressing this limitation in our future work.

**Authors profile**

Agushaka J. O. had his B.Sc. (mathematics with computer science) in 2005 and M.Sc.(computer science) in 2010 all at Ahmadu Bello University, Zaria-Nigeria. He currently lectures at the same institution. He has special interest in semantic web services, knowledge representation, software engineering. All other co-authors are his colleagues in department of mathematics of the same institution.